# Atomically wired molecular junctions: Connecting a single organic molecule by chains of metal atoms

Tamar Yelin [1]†, Ran Vardimon [1]†, Natalia Kuritz †‡, Richard Korytár §, Alexei Bagrets §∥, Ferdinand Evers §⊥, Leeor Kronik ‡, and Oren Tal *†

† Department of Chemical Physics and ‡ Department of Materials and Interfaces, Weizmann Institute of Science, Rehovot, 76100 Israel

§ Institute of Nanotechnology, ∥ Steinbuch Centre for Computing, and ⊥Institut Für Theorie der Kondensierten Materie, Karlsruhe Institute of Technology (KIT), 76128 Karlsruhe, Germany

Email: oren.tal@weizmann.ac.il

**Abstract**

Using a break junction technique, we find a clear signature for the formation of conducting hybrid junctions composed of a single organic molecule (benzene, naphthalene or anthracene) connected to chains of platinum atoms. The hybrid junctions exhibit metallic-like conductance (~0.1-1$G_0$), which is rather insensitive to further elongation by additional atoms. At low bias voltage the hybrid junctions can be elongated significantly beyond the length of the bare atomic chains. Ab initio calculations reveal that benzene based hybrid junctions have a significant binding energy and high structural flexibility that may contribute to the survival of the hybrid junction during the elongation process. The fabrication of hybrid junctions opens the way for combining the different properties of atomic chains and organic molecules to realize a new class of atomic scale interfaces.

---

[1] These authors contributed equally to this work.

Acquiring control over electronic transport at the atomic scale requires the ability to fabricate electronic devices in atomic resolution. The high control over the atomic structure of organic molecules promoted the use of molecular junctions as a test-bed for electronic transport at the atomic scale[1,2,3,4,5,6]. However, the structure of these junctions is limited to a molecule sandwiched between macro-scale electrodes with typical conductance restricted to off-resonance tunneling (several orders of magnitude lower than $G_0$). Ideally, one would like to have the freedom to wire individual organic molecules to other atomic-scale conductors[7,8,9] with minimal conductance attenuation in order to form efficient molecular junctions based on more than a single atomic-scale component. Here we show that benzene and larger organic molecules can be electrically connected by the smallest possible conducting wire: a chain of atoms. The wiring process is done in a mechanically controllable break junction setup by pulling atom after atom from platinum (Pt) electrodes into a molecular junction.

When atomic chains are pulled between metallic electrodes in the presence of diatomic molecules, the chains can be decorated by the small molecules or their decomposed atoms[10,11,12,13,14,15]. In the case of organic molecules connected by thiol groups between two gold electrodes, stretching the junction can lead to its elongation[16,17,18,19]. A direct binding of alkanes to gold electrodes showed similar behavior[20]. In these cases, however, it is not trivial to determine whether the observed elongation is a consequence of atomic chains being pulled from the electrodes or simply a collective deformation of the electrode apexes[16,21]. These junctions are typically characterized by low conductance due to the linking groups at the molecule-electrode interfaces[16,17,18,19,22] or the conductance susceptibility to molecular length[20]. We show that binding organic molecules such as benzene, naphthalene or anthracene directly to Pt electrodes, can form highly conductive hybrid junctions (HJs) with atomic chains bridging the organic molecule and the electrodes. All three HJs have a comparable conductance which is weakly dependent on the chain length. The conditions for the formation of HJs can be optimized to have longer chains than obtained by stretching bare atomic Pt junctions. Ab initio calculations show that wiring benzene with Pt atomic chains leads to a stable molecular junction with significant structural flexibility. Further analysis indicates that the Pt-benzene bond is formed by hybridization between the benzene π-orbitals and the Pt *d*-orbitals.

The HJs were constructed using a mechanically controllable break junction[23] (Fig. 1b, inset) operated at 4.2K. After breaking a Pt wire in cryogenic vacuum to form two ultra-clean electrode tips, the target molecules were introduced via a heated capillary[24] that forms a passage between a molecular source at room temperature and the cold Pt junction. Conductance (I/V) was measured as a function of relative electrode displacement as the junction was broken and reformed repeatedly in order to collect statistical data. Additional experimental details are discussed in the Supplementary Information.

As a first step, benzene was introduced to a bare Pt junction[24]. As a result, a clear change in the junction conductance was observed, indicating the formation of a Pt/benzene junction. Fig. 1e shows several conductance traces measured during breaking of a Pt wire prior to benzene introduction. The last conductance plateau before rupture is attributed to conductance through a cross section of a single atom[25]. The conductance histogram in Fig. 1a is constructed from a set of 10,000 conductance traces. The broad peak at ~1.7$G_0$, where $G_0 = 2e^2/h$ is the conductance quantum, indicates the most probable conductance of a single atom constriction and the tail-shaped distribution at low conductance is the signature of tunneling transport that follows the rupture of atomic contacts[9,26,27]. When benzene is introduced the conductance histogram changes considerably[28,24] (Fig. 1c). The peak at ~1.7$G_0$ vanishes and a new conductance peak appears at ~1.0$G_0$, accompanied by a pronounced tail at lower conductance. This somewhat lower conductance of the molecular junctions is also apparent when comparing individual traces (Fig. 1e,g). Using inelastic electron spectroscopy, the formation of molecular junctions is verified independently, as seen in Supplementary Section 4.

To learn about the structure of the Pt/benzene junctions when the electrode separation is increased, the same set of traces was used to construct a 2D histogram (Fig. 1d). Here the color code represents the number of times that a certain combination of conductance and relative electrode displacement was detected. The observed decrease in conductance from ~1$G_0$ to 0.2$G_0$ as the junction is elongated by ~3Å is in good agreement with previous findings of Kiguchi et al.[28] that attributed this conductance reduction to a transition from a perpendicular orientation to a tilted orientation of the benzene in the junction (Fig. 1d, schemes A and B1, respectively).

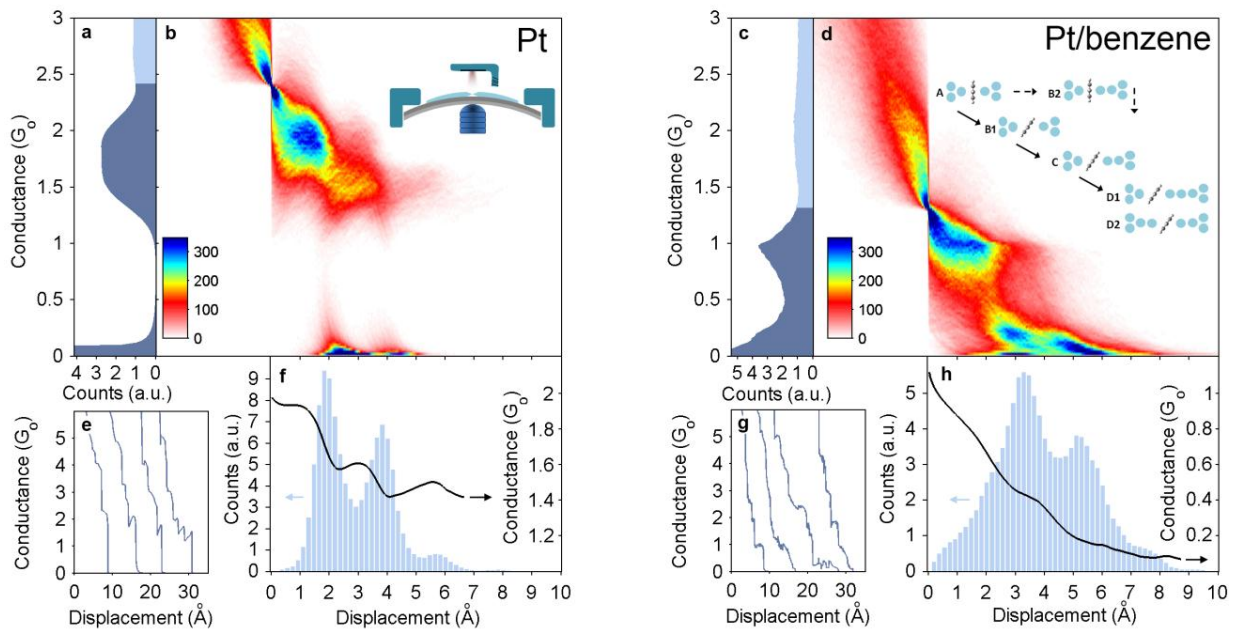

**Figure 1** Statistical analysis of 10,000 conductance traces measured before and after the introduction of benzene to Pt junctions (left and right panel respectively). (a,c) Conductance histograms. The peaks represent the most probable conductance of atomic and single molecule junctions respectively. (b,d) 2D histograms of the number of counts at each conductance-displacement combination. Zero displacement is set to a fixed conductance value (2.4$G_0$ for Pt, 1.3$G_0$ for Pt/benzene) chosen at the top edge of the peak in the conductance histogram. Insets: mechanically controllable break junction apparatus (b) and artist's impression of the evolution of Pt/benzene HJ during stretching (d). (e,g) Examples for conductance traces, shifted for clarity. (f,h) Length histograms (light blue) showing the distribution of junction lengths from zero displacement until rupture (see Supplementary Sections 2,3) and average conductance traces (black curves) showing the evolution of the conductance during stretching. All the measurements presented in this figure were taken at bias voltage of 100mV.

Surprisingly, the Pt/benzene junction is elongated significantly beyond the typical length of an extended Pt-benzene-Pt junction (~3Å elongation in Fig.1d)[28]. To better understand this observation, we studied the distribution of lengths that the junctions can be elongated to. We constructed a length histogram (Fig. 1h) from 10,000 independent molecular junctions (see Supplementary Sections 2,3). The sequence of peaks shows that the junction can have several typical length values, indicating that the junction is elongated each time by a similar repeated unit with a length equal to the distance between the peaks[9]. The average peak separation is 2.0±0.3Å. Interestingly, the peak separation is similar to the typical repeated unit found for Pt atomic chains[9,26]. Smit *et al.*[9] showed that atomic chains can be formed when Pt atomic junctions are elongated. When the distance between the electrodes increases, the junction can either break (revealed as a series of conductance drops in Fig. 1b) or be elongated by pulling an atom from the electrodes. The process can repeat itself to form a chain of several atoms. Fig. 1f shows a length histogram built from conductance traces that were measured for a bare Pt junction right before the introduction of benzene. The sequence of peaks presented here is the signature of Pt atomic chain formation[9] and the average peak separation is 2.0±0.2Å, similar to the value found for the elongated Pt/benzene junction. This indicates that the Pt/benzene junction is most likely elongated by sequential addition of Pt atoms to the molecular junction.

To verify that the repeated unit is indeed a Pt atom, we looked for an additional fingerprint of Pt atomic chains. We compared the average conductance as a function of electrode displacement for sets of traces measured before and after the introduction of benzene (black curves in Fig. 1f,h respectively). The average conductance trace of Pt junctions reveals oscillations with the same peak periodicity in the length histogram[29]. These oscillations were attributed to repeated transitions between a zigzag to a more linear configuration as the chain is elongated[30]. The chain straightening leads to a better orbital overlap which allows for higher conductance. Further stretching eventually reducaes the conductance due to substantial interatomic separation before another atom is pulled in, or due to chain rupture. The average trace of Pt/benzene junctions (Fig. 1h) reveals clear conductance oscillations on top of the overall reduction in conductance. The periodicity of the oscillations is remarkably similar for Pt and Pt/benzene junctions. This additional indication for the presence of Pt atomic chains provides further verification for the formation of hybrids based on molecules and atomic chains. We do not find indications for adsorption of additional molecules on the HJ, such as larger peak separation in the length histogram when molecules are introduced, as previously found for Pt chains exposed to hydrogen[11]. In the latter case, the larger peak separation was ascribed to larger Pt-Pt atomic distance due to chain decoration by hydrogen. Our observation is insensitive to the amount of molecule dosage.

The elongation of benzene HJs can be improved significantly by the right choice of applied voltage across the junction. Fig. 2a shows length histograms of benzene HJs, measured at bias voltage of 50mV and 100mV. The length histogram taken at 50mV reflects a higher probability for the formation of longer HJs while the location of the peaks remains roughly the same. The effect of the applied voltage on the chain length is a general trend as seen in Fig. 2b. Here, the average length of the 10% longest traces is presented as a function of applied voltage. The maximal elongation of bare Pt junctions is only moderately affected by the applied voltage in the presented range. On the other hand, the longest Pt/benzene traces become substantially longer at lower voltage. The different dependence can be understood as a consequence of different sensitivity to heating effects such as Joule heating and voltage excitation of vibrations in the junction,[31–35] as well as current-induced forces [36,37]. The onset of vibration activation of Pt/benzene junctions takes place between 20 and 50 mV[24]. Interestingly, this is the range of voltages in Fig. 2b which exhibits the largest difference in the behavior of the Pt and Pt/benzene junctions. Therefore, the higher sensitivity of Pt/benzene junctions to the applied voltage can be attributed to the activation of molecular junction vibrations. In addition to the stronger length-voltage dependence, at low bias voltage the length of benzene HJs is remarkably higher than the length of bare Pt chains. This exceptional stability of the HJs is not fully understood. However, the richness of possible configurations of HJs compared to the homogeneous Pt chains may play an important role in their remarkable survival under elongation.

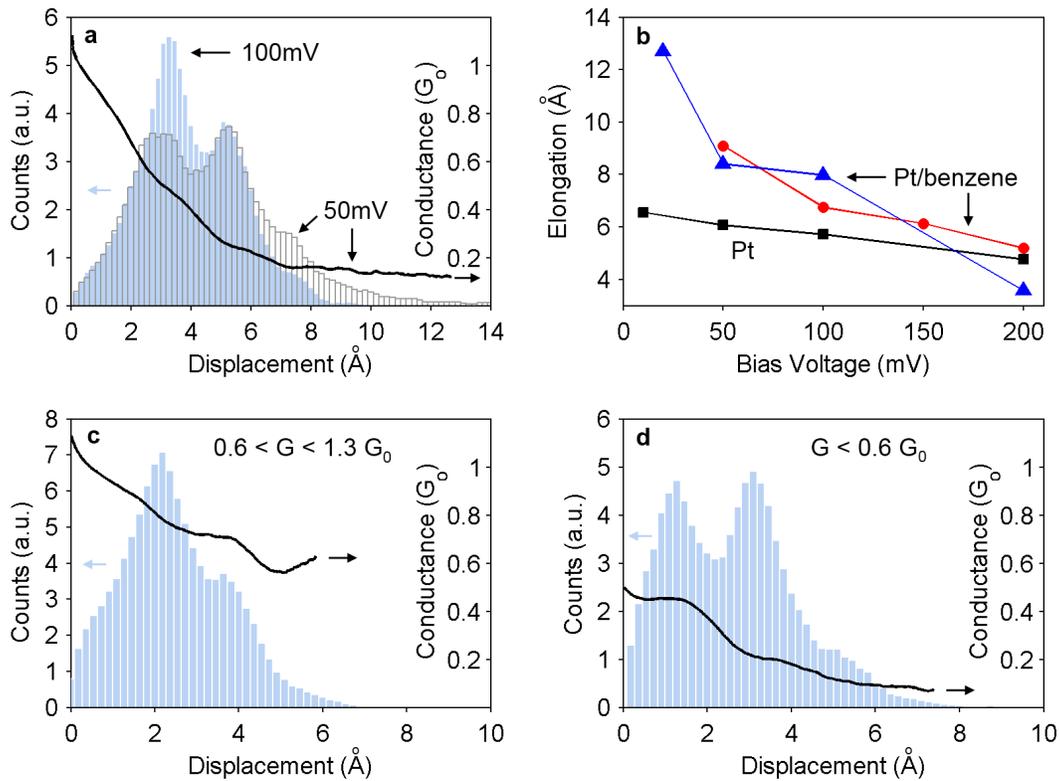

**Figure 2** (a) Length histograms for Pt/benzene junctions based on 10,000 traces each, measured in two sequential measurements at 50mV (dark grey) and 100mV bias voltage (light blue). The length histogram is extended to higher displacement values for the lower bias voltage. The average conductance trace (black curve) is based on measurements at 50mV bias. (b) Length dependence on bias voltage of Pt (squares) and Pt/benzene (circles and triangles) junctions. Each point represents the average length of the 10% longest atomic or molecular junctions and the lines connect sequential points. For Pt/benzene two sequences of measurements are presented. (c,d) Length histograms and average conductance traces measured at 100mV, constructed from the part of each trace inside a conductance window of 1.3-0.6$G_0$ (c) and lower than 0.6$G_0$ (d).

We now examine the evolution of conductance with junction elongation. As mentioned above, when the junction is stretched the conductance decreases from $1G_0$ to $0.2G_0$. This behavior was previously ascribed to molecule tilting[28]. However, in Fig. 1d we can recognize a plateau at ~$1G_0$ which coexists with the general behavior of conductance decrease. Constructing a length histogram and an average trace (Fig. 2c) for the high conductance region ($1.3-0.6G_0$) reveals peaks in the length histogram separated by 1.7±0.3Å and oscillations in the average conductance. These observations indicate that HJs can be formed in a secondary scenario (relevant for ~20% of the traces) with no significant conductance reduction and probably with no substantial molecule tilting (illustrated in Fig. 1d by Scheme B2). Eventually, all traces drop to the low conductance region, below $0.6G_0$, where they can be further elongated (Fig. 2d).

Focusing on the average conductance at high displacement (beyond 5Å in Fig. 2d), a very moderate conductance dependence on elongation is observed. This tendency is better seen for measurements taken at 50mV bias (Fig. 2a, black curve) because the junction elongation is enhanced. Here, beyond the initial conductance reduction, an insignificant conductance dependence on length is clearly observed, indicating a remarkable insensitivity of the average conductance to the introduction of additional Pt atoms.

In order to test whether other organic molecules can form HJs by binding to Pt electrodes, we performed similar experiments with naphthalene and anthracene. Fig. 3 presents length histograms and average traces for Pt/naphthalene and Pt/anthracene molecular junctions. In both cases the junctions are elongated significantly beyond the molecule length (which is roughly 5Å and 7Å respectively[38]). The series of peaks in the length histograms indicates a repeated unit that adds up as the junction is elongated. In comparison to Pt/benzene HJs, the length histograms of naphthalene and anthracene junctions have a higher number of peaks, which indicates that HJs based on these molecules can form hybrids with longer atomic chains. The larger average peak separation found for these junctions (2.4±0.3 and 2.2±0.3Å respectively) can result from more effective rearrangement (e.g. tilting or sliding) of the longer molecules that permits an extended elongation before rupture or insertion of an additional atom. Conductance oscillations can be seen in both cases, although they are more pronounced for Pt/naphthalene. These findings clearly indicate that the formation of stable HJs is possible for different organic molecules and should be further tested as a more general method for wiring organic molecules with atomic chains.

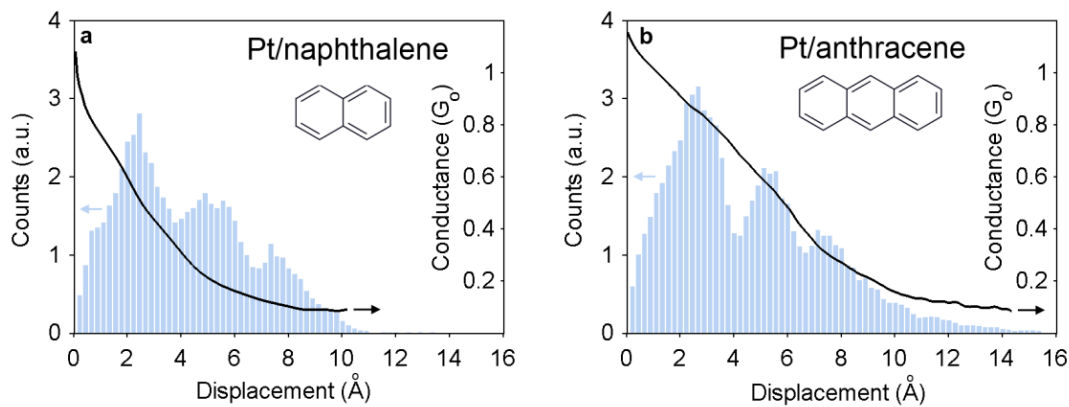

**Figure 3** (a) Length histograms (light blue) and average conductance traces (black curves) of Pt/naphthalene (a) and Pt/anthracene (b) junctions, both measured at 200mV bias voltage and based on more than 5,000 traces.

To better understand the structure and stability of benzene-based HJs, we performed DFT calculations using the Quantum Espresso[39] package (see Supplementary Section 6). We note that our calculations are at equilibrium and therefore do not address the effect of bias. Fig. 4a shows two optimized configurations of a benzene junction with no chain atoms. The benzene tilt is seen when comparing the most stable configuration (M1) and a stretched configuration before breaking (M2). Fig. 4d shows the total energy change as the electrodes are pulled apart. The binding energy of the junction, determined by the difference between the total energy of the most stable configuration and of the ruptured junction, is 1.7±0.1eV. This value indicates significantly stronger binding than a typical physical adsorption, in agreement with previous findings[28]. Note that the calculated Pt-C bond length is about 2.1Å, typical to chemical binding. In Fig. 4e the total energy for symmetric and asymmetric HJs with two chain atoms is plotted versus electrode displacement (sample configurations shown in Fig. 4b,c). For the asymmetric HJ the energy does not increase substantially beyond the most stable configuration due to a low barrier transition to a more compact configuration. The better energetic stability of the asymmetric junction with respect to the symmetric one is manifested in the binding energies: 2.0±0.1eV and 1.3eV±0.1eV, respectively. Interestingly, inelastic electron spectroscopy (see Supplementary Sections 4) is consistent with asymmetric coupling to the electrodes.

The stability of benzene HJs during stretching is facilitated by the additional degrees of freedom in which the HJ can be modified to survive elongation. This is illustrated in Fig. 4f,g for the asymmetric structure. The length of the bond prone to rupture between the molecule and the chain ($d_1$), as well as the angle between this bond and the molecule (α) are only moderately increased for a substantial stretching distance (up to point A2) while the chain becomes more linear and the angle between the chain and the electrode axis (β) is considerably modified. The adjustment of the junction before point A2 also involves moderate changes in the Pt-Pt bonds (e.g. $d_2$). This tendency is inverted beyond A2, where further stretching leads to an increase of α and elongation of $d_1$ until its full rupture. The richness of possible HJ configurations also allows different conformations for similar electrode distance, which are similar in energy (Supplementary Section 7). Such flexibility can lower the probability for junction rupture during elongation.

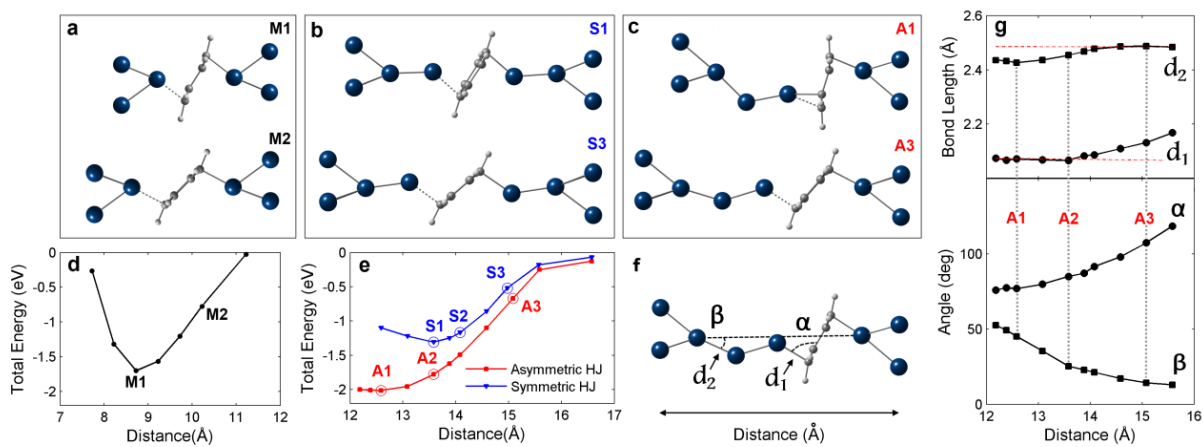

**Figure 4** (a,b,c) Optimized geometries of Pt/benzene junctions calculated for different electrode distances to illustrate possible snapshots during junction stretching. Three possible cases are treated: molecular junction with no additional atoms (a), symmetric HJ with one additional atom at each side (b), asymmetric HJ with two additional atoms at one side (c). For each case, the most stable and a stretched configuration are shown (top and bottom, respectively). The structures were obtained by fixing the distance between the electrodes while optimizing the positions of the molecule, the additional atoms, and the atom at the apex of each electrode. (d,e) Total energy plots showing the calculated total energy for the relaxed structures at each inter-electrode distance for the molecular junction (d) and the symmetric and asymmetric HJs (e). The total energies are presented relative to the corresponding ruptured junction, obtained by splitting the junctions at the bond prone to rupture (according to force analysis), marked in a,b,c by dashed lines. (f) Definitions of relevant bond lengths and angles of the asymmetric HJ (shown in c). (g) Evolution of the defined bond lengths and angles of the asymmetric HJ as a function of inter-electrode distance.

To gain insight into the nature of the bond between the benzene and the frontier Pt atoms, we examined the changes in the bond length and angles of benzene in the junction (Fig. 4a, M1), with respect to an isolated benzene molecule. In the junction, the C-C bonds of the benzene elongate by ~3% and the hydrogen atoms are shifted from the benzene plane by up to 12°. These changes imply that the carbons are gaining some $sp^3$ character, instead of the pure $sp^2$ character in isolated benzene, by creating a new bond with the Pt atom.

Further calculations using the FHI-aims package[40] were performed in order to learn about the orbitals participating in the benzene–Pt bond (Supplementary Section 6). Examining the relative contributions of Pt and benzene orbitals to the Kohn-Sham (KS) eigenstates of a cluster containing benzene in between two Pt electrodes allows for determining their importance to the bond. Two configurations were examined: without additional atoms (similar to M1 in Fig. 4a) and with two additional atoms on one side (similar to A1 in Fig. 4c). In both cases, the cluster eigenstates (e.g Fig. 5), when projected on Pt apex atoms, have a strong $d$ character, among which the contribution of $d_{z^2}$ is usually less significant. $s$ and $p$ orbital contributions were also smaller than that of the $d$ orbitals. This is in contrast to the case of Pt monoatomic chains, where $s$ and $d_{z^2}$ are the dominant orbitals in the Pt-Pt bond[41]. Considering the benzene contribution, the four valence π orbitals (two HOMOs and two LUMOs) play a significantly stronger role than the other orbitals. Thus, the Pt-benzene bond is mainly based on $d$-π hybridization, as can also be inferred by inspection of the eigenstates presented in Fig. 5. Furthermore, most of the eigenstates are spread over both the benzene and the electrodes, revealing electron delocalization over the whole structure. The strong Pt-benzene hybridization is also apparent from the mixing of both HOMO and LUMO π orbitals of the benzene into individual eigenstates (e.g. Fig. 5b).

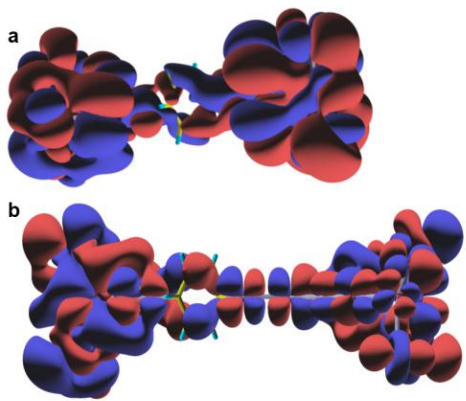

**Figure 5** Examples of calculated KS eigenstates of Pt/benzene clusters corresponding to two types of Pt/benzene junctions: without additional atoms (a) and with two additional atoms on one side of the benzene (b). The presented eigenstates have energies of ~0.01eV & ~0.125eV below the HOMO, respectively. These representative eigenstates are delocalized over the whole Pt/benzene structure. Both of them arise from hybridization mainly between the π antisymmetric HOMO of the benzene with the $d_{xy}$ orbital of the Pt apex atoms. In b there is also some contribution from the π antisymmetric LUMO of the benzene.

To conclude, we have found that benzene and other organic molecules can be electrically wired by chains of Pt atoms, while maintaining high conductance that is insensitive to additional chain elongation once the hybrid is formed. The length of HJs based on benzene depends on the applied voltage and, at sufficiently low voltage, it can exceed the length of pure Pt atomic chains formed at the same conditions. Calculations show that the benzene based HJ is energetically stable and has considerable structural freedom with some energetic preference to HJs with a single chain. The studied HJs demonstrate how individual organic molecules can be wired by the smallest available conducting wire. This is a further step toward more versatile atomic-scale electronic conductors that combine the different worlds of metallic and molecular nanostructures. The HJs form a fascinating class of heterogeneous interfaces between two different low dimensional systems that can serve as an attractive playground for electronic and heat transport at the atomic scale.

**Acknowledgment**

OT acknowledges support of the Israel Science Foundation Grant No. 1313/10, the German-Israeli Foundation Grant No. I-2237-2048.14/2009 (GIF Young), Minerva Foundation and the Herold Perlman Family. LK acknowledges the Israel Science Foundation and the Lise Meitner Minerva Center for Computational Chemistry. FE acknowledges funding by the DFG-Center of Functional Nanostructures and by the DFG Priority Program 1243; AB acknowledges funding by the DFG (research grant BA 4265/2-1). Furthermore, FE and AB thank I. Kondov (SCC) for support and the Juelich Supercomputer Center (project HKA12) for allocation of computing time on JUROPA. Finally, FE also expresses his gratitude to the IAS of Hebrew-University, where part of the work has been performed in the workshop Molecular Electronics, for its warm hospitality.

**References**

(1) Nitzan, A.; Ratner, M. A. *Science* **2003**, *300*, 1384 –1389.

(2) Elbing, M.; Ochs, R.; Koentopp, M.; Fischer, M.; Von Hänisch, C.; Weigend, F.; Evers, F.; Weber, H. B.; Mayor, M. *Proc. Natl. Acad. Sci. U. S. A.* **2005**, *102*, 8815.

(3) Yu, L.; Keane, Z.; Ciszek, J.; Cheng, L.; Tour, J.; Baruah, T.; Pederson, M.; Natelson, D. *Phys. Rev. Lett.* **2005**, *95*, 256803.

(4) Chen, F.; Li, X.; Hihath, J.; Huang, Z.; Tao, N. *J. Am. Chem. Soc.* **2006**, *128*, 15874–15881.

(5) Martin, C. A.; Ding, D.; Sørensen, J. K.; Bjørnholm, T.; van Ruitenbeek, J. M.; van der Zant, H. S. J. *J. Am. Chem. Soc.* **2008**, *130*, 13198–13199.

(6) Venkataraman, L.; Klare, J. E.; Tam, I. W.; Nuckolls, C.; Hybertsen, M. S.; Steigerwald, M. L. *Nano Lett.* **2006**, *6*, 458–462.

(7) Yanson, A. I.; Bollinger, G. R.; van den Brom, H. E.; Agrait, N.; van Ruitenbeek, J. M. *Nature* **1998**, *395*, 783–785.

(8) Ohnishi, H.; Kondo, Y.; Takayanagi, K. *Nature* **1998**, *395*, 780–783.

(9) Smit, R.; Untiedt, C.; Yanson, A.; van Ruitenbeek, J. *Phys. Rev. Lett.* **2001**, *87*, 266102.

(10) Csonka, S.; Halbritter, A.; Mihály, G. *Phys. Rev. B* **2006**, *73*, 075405.

(11) Kiguchi, M.; Stadler, R.; Kristensen, I.; Djukic, D.; van Ruitenbeek, J. *Phys. Rev. Lett.* **2007**, *98*, 146802.

(12) Thijssen, W. H. A.; Strange, M.; van Ruitenbeek, J. M.; others *New J. Phys.* **2008**, *10*, 033005.

(13) Nakazumi, T.; Kiguchi, M. *J. Phys. Chem. Lett.* **2010**, *1*, 923–926.

(14) Kiguchi, M.; Hashimoto, K.; Ono, Y.; Taketsugu, T.; Murakoshi, K. *Phys. Rev. B* **2010**, *81*, 195401.

(15) Makk, P.; Balogh, Z.; Csonka, S.; Halbritter, A. *Nanoscale* **2012**, *4*, 4739–4745.

(16) Wu, S.; Gonzalez, M. T.; Huber, R.; Grunder, S.; Mayor, M.; Schonenberger, C.; Calame, M. *Nat. Nanotechnol.* **2008**, *3*, 569–574.

(17) Xiao, X.; Xu, B.; Tao, N. J. *Nano Lett.* **2004**, 4, 267–271.

(18) Kim, Y.; Hellmuth, T. J.; Bürkle, M.; Pauly, F.; Scheer, E. *ACS Nano* **2011**, *5*, 4104–4111.

(19) Kim, Y.; Song, H.; Strigl, F.; Pernau, H.-F.; Lee, T.; Scheer, E. *Phys. Rev. Lett.* **2011**, *106*, 196804.

(20) Cheng, Z.-L.; Skouta, R.; Vazquez, H.; Widawsky, J. R.; Schneebeli, S.; Chen, W.; Hybertsen, M. S.; Breslow, R.; Venkataraman, L. *Nat. Nanotechnol.* **2011**, 6, 353–357.

(21) Huisman, E. H.; Trouwborst, M. L.; Bakker, F. L.; de Boer, B.; van Wees, B. J.; van der Molen, S. J. *Nano Lett.* **2008**, *8*, 3381–3385.

(22) Ferrer, J.; García-Suárez, V. *Phys. Rev. B* **2009**, *80*, 085426.

(23) Muller, C. J.; van Ruitenbeek, J. M.; de Jongh, L. J. *Phys. C (Amsterdam, Neth.)* **1992**, *191*, 485–504.

(24) Tal, O.; Kiguchi, M.; Thijssen, W.; Djukic, D.; Untiedt, C.; Smit, R.; van Ruitenbeek, J. *Phys. Rev. B* **2009**, *80*, 085427.

(25) Agraıt, N.; Yeyati, A. L.; van Ruitenbeek, J. M. *Phys. Rep.* **2003**, *377*, 81–279.

(26) Smit, R.; Untiedt, C.; Rubio-Bollinger, G.; Segers, R.; van Ruitenbeek, J. *Phys. Rev. Lett.* **2003**, *91*, 076805.

(27) Smit, R. H. M.; Noat, Y.; Untiedt, C.; Lang, N. D.; van Hemert, M. C.; van Ruitenbeek, J. M. *Nature* **2002**, *419*, 906–909.

(28) Kiguchi, M.; Tal, O.; Wohlthat, S.; Pauly, F.; Krieger, M.; Djukic, D.; Cuevas, J.; van Ruitenbeek, J. *Phys. Rev. Lett.* **2008**, *101*, 046801.

(29) Shiota, T.; Mares, A.; Valkering, A.; Oosterkamp, T.; van Ruitenbeek, J. M. *Phys. Rev. B* **2008**, *77*, 125411.

(30) García-Suárez, V.; Rocha, A.; Bailey, S.; Lambert, C.; Sanvito, S.; Ferrer, J. *Phys. Rev. Lett.* **2005**, *95*, 256804.

(31) Horsfield, A. P.; Bowler, D. R.; Ness, H.; Sánchez, C. G.; Todorov, T. N.; Fisher, A. J. *Rep. Prog. Phys.* **2006**, *69*, 1195.

(32) Avouris, P. *Acc. Chem. Res.* **1995**, *28*, 95–102.

(33) Smit, R. H. M.; Untiedt, C.; van Ruitenbeek, J. M. *Nanotechnology* **2004**, *15*, S472.

(34) Huang, Z.; Xu, B.; Chen, Y.; Di Ventra, M.; Tao, N. *Nano Lett.* 2006, 6, 1240–1244.

(35) Stipe, B. C.; Rezaei, M. A.; Ho, W.; Gao, S.; Persson, M.; Lundqvist, B. I. *Phys. Rev. Lett.* **1997**, 78, 4410–4413.

(36) Dundas, D.; McEniry, E. J.; Todorov, T. N. *Nat. Nanotechnol.* **2009**, *4*, 99–102.

(37) Lü, J.-T.; Brandbyge, M.; Hedegård, P. *Nano Lett.* **2010**, *10*, 1657–1663.

(38) Cruickshank, D. W. J.; Sparks, R. A. *Proc. R. Soc. London, Ser. A* **1960**, *258*, 270 –285.

(39) Giannozzi, P.; Baroni, S.; Bonini, N.; Calandra, M.; Car, R.; Cavazzoni, C.; Ceresoli, D.; Chiarotti, G. L.; Cococcioni, M.; Dabo, I.; Dal Corso, A.; de Gironcoli, S.; Fabris, S.; Fratesi, G.; Gebauer, R.; Gerstmann, U.; Gougoussis, C.; Kokalj, A.; Lazzeri, M.; Martin-Samos, L.; Marzari, N.; Mauri, F.; Mazzarello, R.; Paolini, S.; Pasquarello, A.; Paulatto, L.; Sbraccia, C.; Scandolo, S.; Sclauzero, G.; Seitsonen, A. P.; Smogunov, A.; Umari, P.; Wentzcovitch, R. M. *J. Phys.: Condens. Matter* **2009**, *21*, 395502.

(40) Blum, V.; Gehrke, R.; Hanke, F.; Havu, P.; Havu, V.; Ren, X.; Reuter, K.; Scheffler, M. *Comput. Phys. Commun.* **2009**, *180*, 2175–2196.

(41) Nielsen, S.; Brandbyge, M.; Hansen, K.; Stokbro, K.; van Ruitenbeek, J.; Besenbacher, F. *Phys. Rev. Lett.* **2002**, *89*.

# Supporting Information

**Contents**

1. Experimental techniques
2. Length calibration
3. Identifying the transition to the tunneling regime
4. Asymmetric coupling observed by inelastic electron spectroscopy
5. Bias dependence of Pt/benzene junctions
6. DFT calculations
7. Energetically close-lying configurations of the HJ
8. References

**1. Experimental techniques**

Our experiments are conducted using a mechanical controllable break junction (MCBJ) setup at 4.2K. The sample is fabricated by attaching a notched Pt wire (99.99%, 0.1mm diameter, Goodfellow) to a flexible substrate (1mm thick phosphor bronze covered by 100μm insulating Kapton film). The substrate is bent in cryogenic vacuum in order to break the wire and form an adjustable gap between two ultra-clean atomically-sharp tips. A piezoelectric element (PI P-882 PICMA) is used to tune the bending of the substrate and control the distance between the electrodes with sub-Angstrom resolution. The piezoelectric element is driven by a 16 bit NI-PCI6221 DAQ card connected to a high peak current piezo driver (Piezomechanik SVR 150/1). The Pt wire is repeatedly pulled apart till rupture and reformed to about $50G_0$ while the conductance is measured simultaneously. The conductance traces vs. the voltage applied to the piezoelectric element are collected at a rate of ~20Hz with a sampling rate of 100kHz. In order to measure the conductance, a constant bias voltage is given by the DAQ card and divided by 10 to lower the noise. The current is amplified by a current preamplifier (SR570) and then measured by the DAQ card. Inelastic electronic spectroscopy (IES) measurements, in which differential conductance (dI/dV) is measured as function of bias voltage, were performed using a lock-in amplifier (SR830). The differential conductance is obtained from the ac response to a small reference signal (1mVrms, ~4kHz) added to the bias voltage as the bias is swept from 100 to -100mV and back.

The different molecules used, namely: benzene (99.97% purity), naphthalene (99.9%) and anthracene (99.9%), all purchased from Sigma Aldrich, were introduced into the cryogenic environment via a heated capillary[1]. Before each experimental session the capillary was baked out overnight at ~150ºC. Prior to molecule insertion, benzene was degassed by several freeze-pump-thaw cycles. In the case of naphthalene and anthracene the degassing was done by three pumping cycles. Because of the temperature gradient along the capillary and possible shading by the Pt electrode geometry, it is hard to determine the molecule coverage in the junction vicinity. Before inserting molecules, several sets of 5000 conductance traces or more were collected to verify that the conductance histograms show the signature of a clean Pt junction. After the molecules were introduces, a significant change in the conductance histogram was recorded, indicating the formation of a molecular junction.

## 2. Length calibration

The MCBJ setup enables the control of the relative electrode displacement with sub-Angstrom accuracy owing to its three point bending mechanism. The whole range of elongation for a typical trace is in the order of tens of Angstroms. In this range the displacement ratio $k$ between the electrode separation $\delta$ and the voltage applied on the piezoelectric element $V_p$ (i.e. $k = \delta/V_p$) is approximately constant. In order to determine the displacement ratio we use the procedure of Untiedt *et al.*[2] that assumes an exponential decay of conductance at the tunneling regime. For a tunneling gap between the Pt electrodes, the conductance dependence on the electrode separation is fitted to the exponential decay expected from a rectangular barrier:

$$R \propto exp\left(\frac{2}{\hbar}\sqrt{2m\phi}\delta\right)$$

where R is the measured resistance, m is the electron mass, $\phi$ is the work function taken to be 5.7eV for Pt[3] and $\hbar$ is the reduced Planck constant. The displacement ratio can be calculated from the slope of $lnR$ $(V_p)$ using the following relation:

$$k = \frac{\hbar}{2\sqrt{2m\phi}}\frac{\partial(lnR)}{\partial V_p}$$

We applied this analysis on 8 different sets of traces, measured on bare Pt junctions, each consists at least 5000 conductance traces, measured at different values of bias voltage. For each set, the displacement ratio was determined by the median of the $k$ values obtained from the traces. We typically found a standard deviation of ~30% in $k$ which can be accounted to the uncertainty in the work function of the Pt apexes and in the shape of the tunneling barrier. Nevertheless, the median $k$ in our measurements maintained a value of $91\text{Å}/V \pm 4\%$ between the different sets of traces.

A verification of our length calibration was done by analyzing the length histograms of the last Pt plateau (e.g. Fig. 1f). The peaks in the histogram appear at preferred length values for atomic chain rupture, which have a length periodicity given by the contribution of an added Pt atom[4]. We calculate the average inter-peak separation between the first 3 peaks of such histograms using the above displacement ratio. This yields a value of 2.0±0.2Å, which is invariable between the different trace sets. This value is highly reproducible and by independently calibrating the displacement ratios of different samples by the above mentioned procedure, we get inter-peak displacement values that vary within the range of error. Our results are in agreement with the value of 2.3±0.2Å obtained by Smit *et al.*[4]

## 3. Identifying the transition to the tunneling regime

In order to produce a reliable length histogram it is essential to have a procedure for precisely determining the start and end points of the region identified as the molecular trace. In "molecular trace" we refer to the part of the conductance trace in which the signature of the molecule is apparent as conductance plateaus with values that do not appear for bare Pt (usually below $1.3G_0$). The typical procedure used in previous works is based on setting conductance limits that bound the peak of interest in the conductance histogram[4,5]. This procedure works well when there is a well defined peak in the conductance histogram such as the peak related to conductance through a cross-section of a single Pt atom. However, this is not the case for molecular junctions as Pt/benzene since the low conductance values of the molecular junction are not well separated from conductance values of vacuum tunneling following rupture.

The difficulty is illustrated in several example traces measured on a Pt/benzene junction and presented in Fig. S1a. The beginning of the molecular trace is set to the first data point with a conductance bellow $1.3G_0$ (marked with a green horizontal line), which is an upper bound for the $\sim 1G_0$ peak in the conductance histogram (see Fig. 1c). The point of rupture however cannot be determined precisely by setting a fixed conductance value. For example, using a lower bound of $0.1G_0$ (upper red line) will miss plateaus with low conductance, such as in the case of the two rightmost traces. On the other hand, using a lower bound of $0.01G_0$ (lower red line) will take into account the contribution of through-vacuum tunneling as in the case of the two leftmost traces. Such contributions can erroneously add up to 2Å to the length of the molecular trace, making the length analysis less accurate.

To overcome this setback we introduced a new procedure for identifying the rupture of the molecular junction based on detecting discontinuities in the conductance. This method takes advantage of the fact that the onset of tunneling is identified by an abrupt reduction in the conductance. The procedure goes as follows: (1) divide the trace into continuous segments separated by a large discontinuity; (2) join consecutive segments with conductance difference smaller than a certain value; (3) the onset of tunneling is the first point of the last segment. Here, step (1) divides the trace by jumps in the conductance while step (2) is used to ignore discontinuities caused by noise. Since tunneling is not characterized by conductance discontinuities, only in rare cases the last segment does not include the whole tunneling region. An important advantage is that in the case this procedure errs, the error will result in a slightly shorter molecular trace. This ensures that the extensive elongations observed for Pt/benzene junctions are not artifacts which could occur by using a conductance bound in the tunneling range. To confirm the validity of this method, hundreds of traces were manually checked for the three types of molecular junctions presented here (Pt/benzene, Pt/naphthalene and Pt/anthracene) and the percentage of errors was found to be lower than 5% (each error leads to an underestimated length value). Fig. S1a illustrates how the new procedure can overcome the difficulty of distinguishing the rupture of the trace, where simple cut at a certain conductance value fails.

Finally, in Fig. S1b we compare the length histograms produced using the tunneling identification procedure to histograms using different low conductance limits ($0.1G_0$ and $0.01G_0$). Aside from some small differences in the long tail distribution, the series of peaks is similar in all cases. However, the average length identified by our procedure is approximately 10% longer than the 0.1$G_0$ histogram and 10% shorter than the 0.01$G_0$, indicating that the systematic errors given by applying an arbitrary low conductance limit give rise to a total shift of the length histogram although the peak shape is maintained.

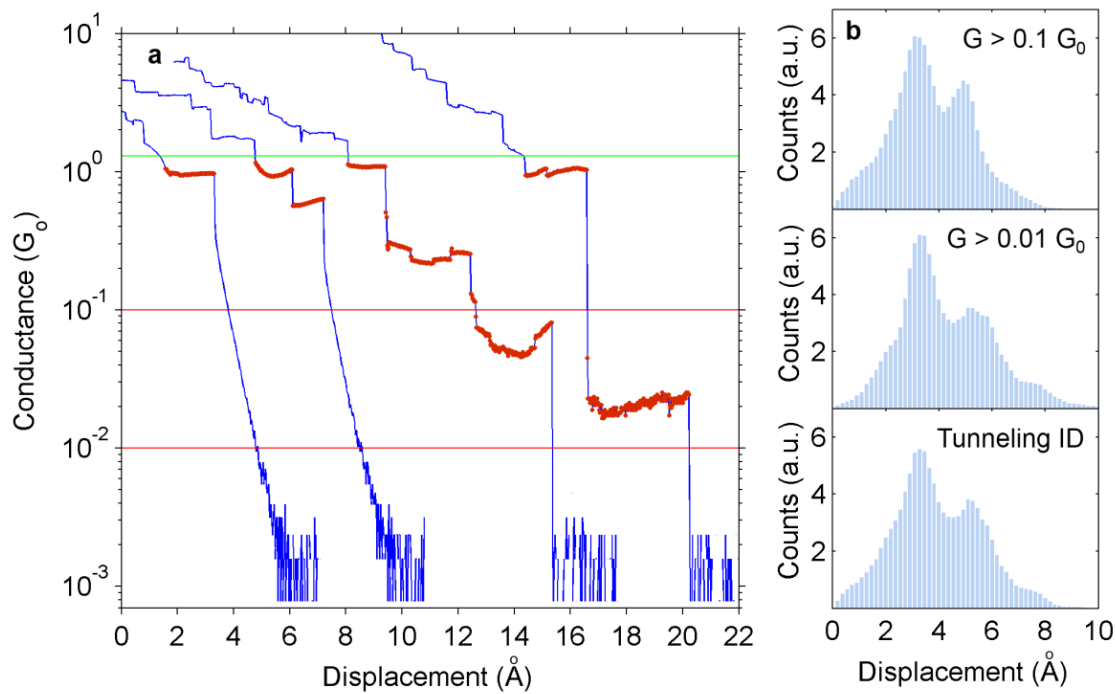

**Figure S1: a.** Conductance traces of Pt/benzene shown in a logarithmic scale. Red dots show the identified plateaus below 1.3$G_0$ (green horizontal line) until the rupture point determined by the tunneling identification procedure (see text). Red lines at 0.1, 0.01$G_0$ are examples for defining the end of the molecular trace using a fixed conductance value. **b.** Pt/benzene length histograms in which the rupture point is defined by 0.1$G_0$ (top), 0.01$G_0$ (middle) or by tunneling identification (bottom).

## 4. Asymmetric coupling observed by inelastic electron spectroscopy

We used inelastic electron spectroscopy (IES) in order to detect the vibration signature of the Pt/benzene junctions using a standard lock-in technique (see section 1). Several examples of IES curves showing typical "step up" (blue) and "step down" (red) features[6] in the conductance are shown in Fig. S2. The step features appear when the applied voltage is equal to the energy needed to activate a molecular vibration by the current carrying electrons ($\hbar\omega = eV$). This interpretation was previously confirmed by measuring the mass dependence of the step voltage using isotope substitution[7]. We performed IES measurements on different Pt/benzene junctions. In order to have independent junction configurations at each measurement, the junction was first rebuilt to have a conductance of ~50$G_0$ and then the electrode displacement was adjusted to stabilize a new junction at a certain conductance in the range of 0.05–1.1$G_0$. We analyzed only spectra exhibiting a clear step behavior observed above 25mV. This lower limit is set to avoid collecting the signature of metal phonons[8]. Clear vibrational signatures were detected in the whole 0.05–1.1$G_0$ range, indicating the formation of molecular junctions with variable characteristic conductance values. The relative step heights (typically 2-10%) are higher than in the case of steps caused by electron-phonon interaction in metallic point contacts[9]. This implies a strong electron-vibration interaction which is typical to a molecular bridge[10].

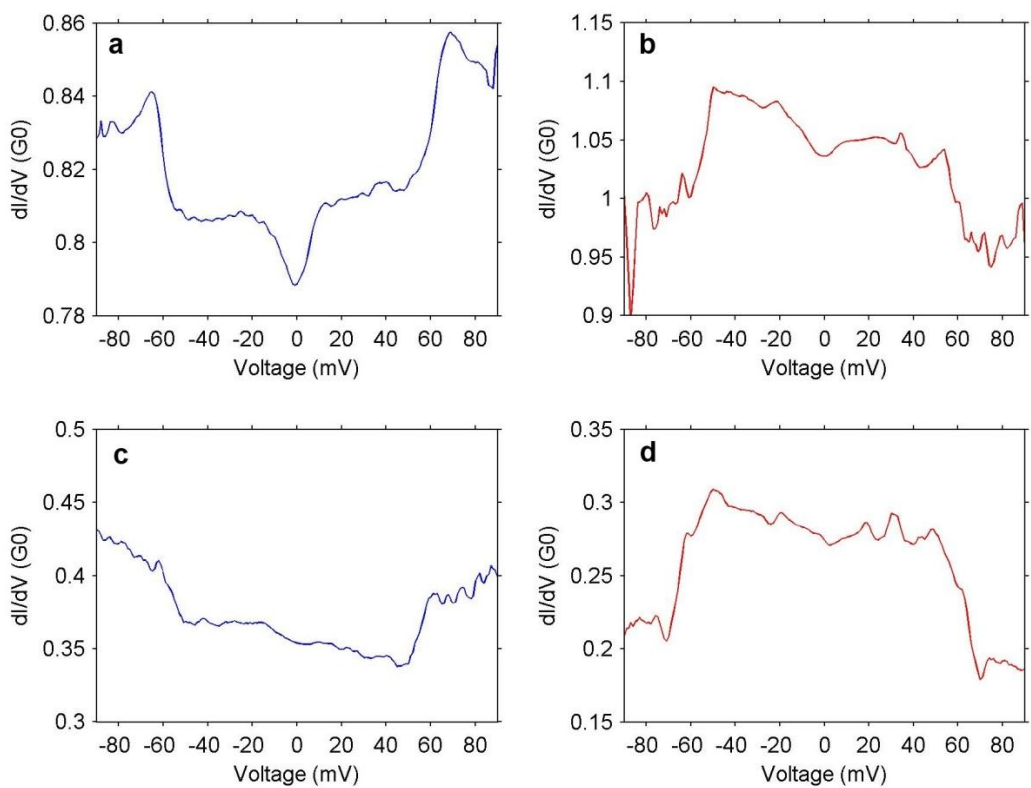

**Figure S2:** IES taken on Pt/benzene junctions with relatively high (**a,b**) and low (**c,d**) conductance. Both "step up" (blue; **a,c**) and "step down" (red; **b,d**) features can appear in a large conductance range.

The step direction (up or down) depends on the transmission of the junction and the symmetry of the coupling to the electrodes[6,11,12]. Under the conditions of our experiment and in the case of a single conduction channel and a symmetric junction the crossover is expected at transmission of $0.5$[13]. However, if the molecule coupling to the leads is asymmetric then the transition value is lowered to $T = 2\alpha/(1+\alpha)^2$, where $\alpha = \Gamma_L/\Gamma_H$ and $\Gamma_H, \Gamma_L$ are the high and low coupling to the different electrodes[11,12]. This equation could be used to estimate the asymmetry in the coupling to the leads. For example, a step down at $0.3 G_0$ implies an asymmetry factor of at least $\alpha = 4$. Note that this argumentation also holds when more than one conduction channel is involved such as in the case of Pt/benzene[7]. However, since in a multi-channel junction the conductance is not equal to the transmission, only a lower bound for the asymmetry can be determined.

To focus on IES that was measured mainly on HJs we limited the analysis to junctions with conductance between $0.05$-$0.3 G_0$. The distribution of step up and step down features is shown in Fig. S3. We have found a clear step down for 20% of the curves indicating a considerable asymmetry in the coupling to the leads ($\alpha \geq 4$). This is a lower limit for the fraction of asymmetric junctions since a step up does not indicate the junction symmetry. While the benzene molecule is highly symmetric, the large asymmetry in the coupling could be explained by the formation of an asymmetric structure of the HJ where the benzene molecule is connected to a Pt chain on one side and attached directly to the electrode apex on the other side.

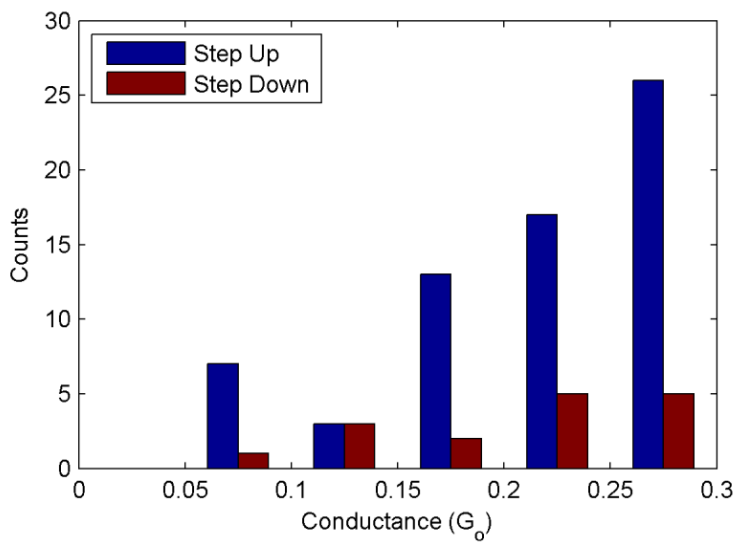

**Figure S3:** Distribution of step up (blue) and step down (dark red) features in IES measured at low conductance values.

## 5. Bias dependence of Pt/benzene junctions

A bias voltage applied over atomic or molecular junctions can lead to local heating caused by activation of phonons[14–17]. In general, at higher effective temperatures the potential barrier for chain rupture can be more easily overcome, leading to increased probability for rupture. Activation of local vibration modes (e.g. in molecular junctions) can assist in specific bond breaking[18] and current-induced non-equilibrium and non-conservative forces can facilitate rupture as well[19,20]. Focusing on the 10% longest junctions in Fig. 2b, the effect of bias voltage on the length of HJ is stronger than for bare Pt atomic chains. In particular, a pronounced decrease in the junction length is observed only for Pt/benzene junctions when the voltage increases from 20mV to 50mV. For these junctions, the excitation of molecular vibrations by conduction electrons is activated exactly in the same voltage range[1]. Thus, the abrupt drop in the length of the HJs may be attributed to instability assisted by activation of vibrations. This possibility is supported by the frequent appearance of noise in the dI/dV spectra of Pt/benzene junctions which is correlated to the onset of vibrations (observed as steps in the dI/dV spectra). For example, Fig. S4 shows that a conductance step in the dI/dV spectra is followed by intensified noise at higher voltages. In these cases, the increased noise indicates that the onset of vibrations results in structural instability.

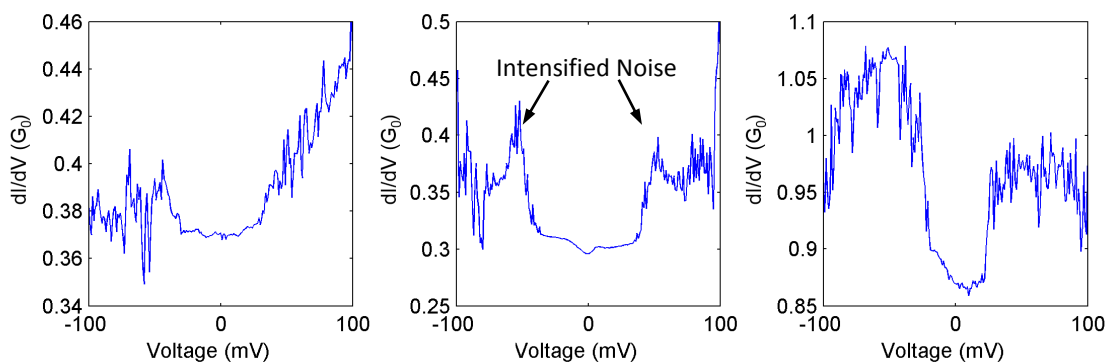

**Figure S4:** dI/dV spectra measured on Pt/benzene junctions.

The effect of applied voltage on the Pt/benzene junction is also observed in conductance histograms (Fig. S5a). As the voltage is increased to 150mV, the total area up to the $1G_0$ peak is gradually reduced indicating less frequent detection of conductance typical to HJs. This is consistent with the lower probability to form long HJs observed in Fig. 2b. Thus at this voltage range molecular junctions are still preserved as indicated by the $1G_0$ peak although the probability to form HJs is reduced.

A more pronounced change in the conductance histogram is observed when the voltage is increased to 200mV. The conductance typical to molecular junctions is strongly suppressed (both the peak at $1G_0$ and the tail at lower conductance). Moreover, a reappearance of the peak at ~$1.7G_0$, which is ascribed to a bare Pt atomic junction, can be observed. These changes indicate that the formation of molecular junctions is significantly inhibited at this voltage and bare Pt junctions are more frequently recovered.

The probability for preservation of molecular junctions can be analyzed by the last conductance value recorded before rupture, $G_{break}$. As a practical approach, traces having $G_{break}<1.3G_0$ are presumed to indicate the formation of a molecular junction, based on the typical conductance of Pt and Pt/benzene junctions (see Fig. 1a and 1c). Figure S5b shows the probability of a measured trace to exhibit a signature of a molecular junction ($G_{break}<1.3G_0$) as a function of bias voltage. For voltage between 20-150mV almost 100% of the traces exhibits the signature of molecular junctions. However, this figure is lowered to about 70% at 200mV, indicating a lower probability to form a molecular junction.

To conclude, the stability of the HJ is significantly reduced at bias voltages corresponding to the onset of molecular junction vibrations (20-50mV). Further increase in voltage (up to 150mV) reduces the HJ length, however, the molecular junction is still preserved. Only above 150mV the probability to form molecular junctions is reduced significantly and a bare Pt atomic junction is more frequently formed.

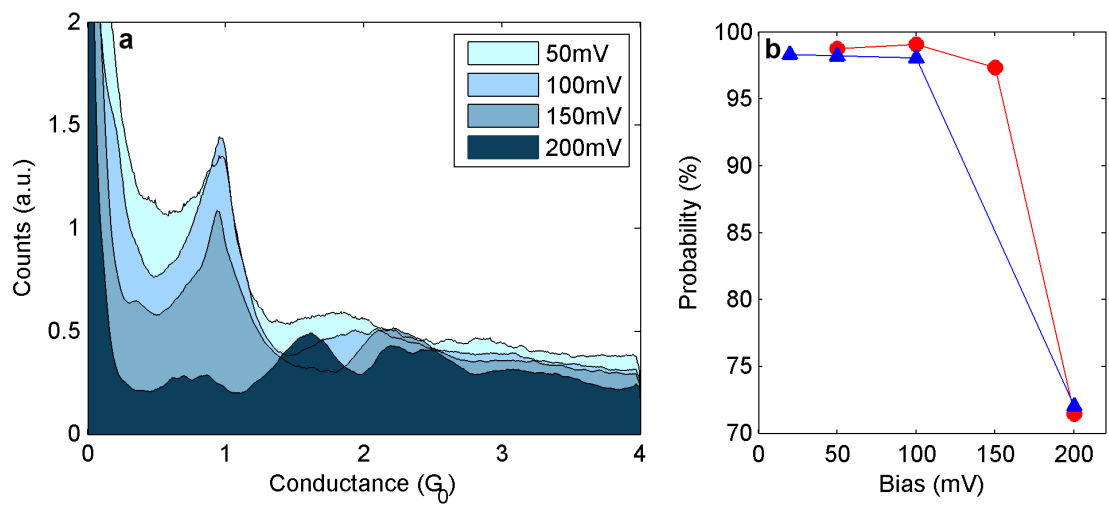

**Figure S5:** (a) Evolution of a Pt/benzene conductance histogram with bias voltage. The histograms consist of 10,000 traces each and are constructed using the same bin size. (b) The probability to measure $G_{break}<1.3G_0$ as function of bias voltage for two different measurement sequences (blue triangles and red circles). The data is shown for the same measurement sequences presented in Fig. 2b. For each bias value, the presented probability is calculated based on at least 5,000 traces.

## 6. DFT calculations

DFT calculations for junction geometries and energies were performed using the Quantum Espresso package[21], which is based on plane waves[22]. We applied fully-relativistic ultra-soft pseudopotentials, as generated by Dal Corso[23] (rrkj3) to describe the Pt atoms. The PBE[24] exchange-correlation functional was used for all atoms. The electrodes were modeled by clusters of 8 Pt atoms, ordered in an FCC lattice (lattice constant 3.92Å[25]) and an apex oriented in the (111) direction[7]. Control calculations using larger electrodes gave similar results for the binding energies. Comparing fully-relativistic and scalar-relativistic calculations showed deviations within the range of computational error, indicating that a scalar-relativistic calculation is sufficient to describe the system within our desired accuracy. The junction configurations were found by fixing the distance between the electrodes and optimizing the position of the last atom at each electrode apex as well as the positions of the chain and the benzene atoms.

Analysis of the Pt/benzene binding was performed using the FHI-aims package[26], with the PBE[24] exchange-correlation functional and a tier2 all-electron numerical basis set with scalar-relativistic corrections for all atoms. The electrodes were modeled by 11 Pt atoms ordered as mentioned above. Both calculation methods gave similar converged structures.

We note that our calculations are at equilibrium and therefore do not address the effect of bias. Nevertheless, as shown below, they do offer insights into the general mechanisms underlying the experimentally observed phenomena.

## 7. Energetically close-lying configurations of the HJ

The more complex structure of the HJs with respect to the initial molecular junction can provide a variety of stable configurations at each electrode distance. For example, Fig. S6 shows two different structures that have very similar total energy and electrode distance. The existence of stable configurations which are energetically close-lying for a given electrode distance increases the probability that the junction would find a stable configuration at each stage of the dynamic process of stretching. This effect can reduce the probability for rupture and promote the HJ elongation.

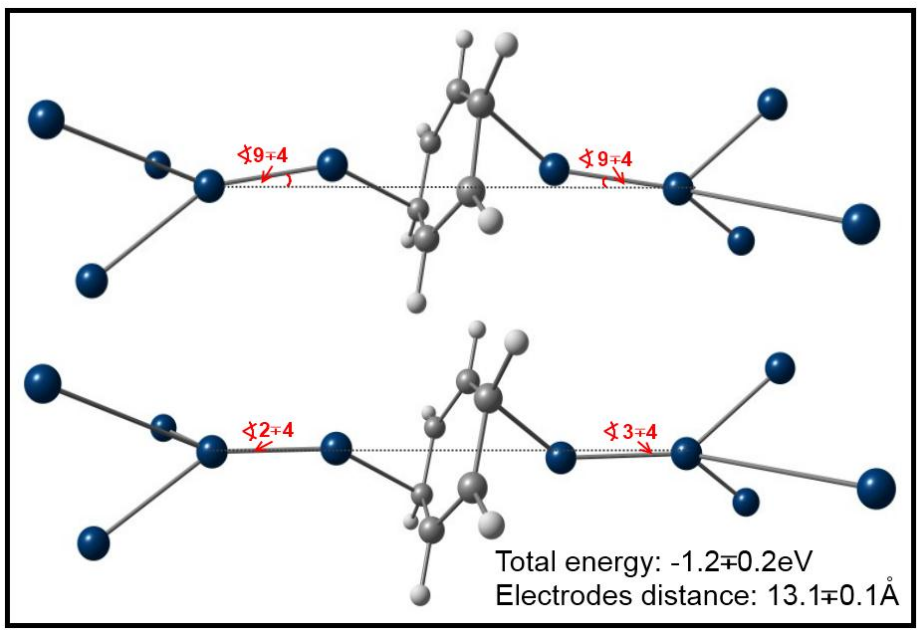

**Figure S6:** An example of two energetically close-lying configurations found for the symmetric HJ for electrode distance of 13.1±0.1Å with a total energy of -1.2±0.2eV. Note that in the upper panel the angles between the ad-atoms and the junction axis are higher, giving it a more "zig-zag" character, compared to the lower panel configuration.

## 8. References

(1) Tal, O.; Kiguchi, M.; Thijssen, W. H. A.; Djukic, D.; Untiedt, C.; Smit, R. H. M.; van Ruitenbeek, J. M. *Phys. Rev. B* **2009**, *80*, 085427.

(2) Untiedt, C.; Yanson, A. I.; Grande, R.; Rubio-Bollinger, G.; Agraït, N.; Vieira, S.; van Ruitenbeek, J. M. *Phys. Rev. B* **2002**, *66*, 085418.

(3) Smit, R. H. M. Ph. D. thesis, Universiteit Leiden, The Netherlands, **2003**.

(4) Smit, R. H. M.; Untiedt, C.; Rubio-Bollinger, G.; Segers, R. C.; van Ruitenbeek, J. M. *Phys. Rev. Lett.* **2003**, *91*, 076805.

(5) Yanson, A. I.; Bollinger, G. R.; van den Brom, H. E.; Agrait, N.; van Ruitenbeek, J. M. *Nature* **1998**, *395*, 783–785.

(6) Tal, O.; Krieger, M.; Leerink, B.; van Ruitenbeek, J. M. *Phys. Rev. Lett.* **2008**, *100*, 196804.

(7) Kiguchi, M.; Tal, O.; Wohlthat, S.; Pauly, F.; Krieger, M.; Djukic, D.; Cuevas, J. C.; van Ruitenbeek, J. M. *Phys. Rev. Lett.* **2008**, *101*, 046801.

(8) Naidyuk, Y. G.; Yanson, I. K. *Point-Contact Spectroscopy*; Springer, 2005.

(9) Agraït, N.; Untiedt, C.; Rubio-Bollinger, G.; Vieira, S. *Chemical Physics* **2002**, *281*, 231–234.

(10) Djukic, D.; Thygesen, K. S.; Untiedt, C.; Smit, R. H. M.; Jacobsen, K. W.; van Ruitenbeek, J. M. *Phys. Rev. B* **2005**, *71*, 161402.

(11) Paulsson, M.; Frederiksen, T.; Ueba, H.; Lorente, N.; Brandbyge, M. *Phys. Rev. Lett.* **2008**, *100*, 226604.

(12) Kim, Y.; Pietsch, T.; Erbe, A.; Belzig, W.; Scheer, E. *Nano Lett.* **2011**, 11, 3734–3738.

(13) Cuevas, J. C.; Scheer, E. *Molecular Electronics: An Introduction to Theory and Experiment*; World Scientific Pub Co Inc, **2010**.

(14) Horsfield, A. P.; Bowler, D. R.; Ness, H.; Sánchez, C. G.; Todorov, T. N.; Fisher, A. J. *Rep. Prog. Phys.* **2006**, *69*, 1195.

(15) Avouris, P. *Acc. Chem. Res.* **1995**, *28*, 95–102.

(16) Smit, R. H. M.; Untiedt, C.; van Ruitenbeek, J. M. *Nanotechnology* **2004**, *15*, S472.

(17) Huang, Z.; Xu, B.; Chen, Y.; Di Ventra, M.; Tao, N. *Nano Lett.* **2006**, *6*, 1240–1244.

(18) Stipe, B. C.; Rezaei, M. A.; Ho, W.; Gao, S.; Persson, M.; Lundqvist, B. I. *Phys. Rev. Lett.* **1997**, 78, 4410–4413.

(19) Dundas, D.; McEniry, E. J.; Todorov, T. N. *Nat Nano* **2009**, *4*, 99–102.

(20) Lü, J.-T.; Brandbyge, M.; Hedegård, P. *Nano Lett.* **2010**, *10*, 1657–1663.

(21) Giannozzi, P.; Baroni, S.; Bonini, N.; Calandra, M.; Car, R.; Cavazzoni, C.; Ceresoli, D.; Chiarotti, G. L.; Cococcioni, M.; Dabo, I.; Dal Corso, A.; De Gironcoli, S.; Fabris, S.; Fratesi, G.; Gebauer, R.; Gerstmann, U.; Gougoussis, C.; Kokalj, A.; Lazzeri, M.; Martin-Samos, L.; Marzari, N.; Mauri, F.; Mazzarello, R.; Paolini, S.; Pasquarello, A.; Paulatto, L.; Sbraccia, C.; Scandolo, S.; Sclauzero, G.; Seitsonen, A. P.; Smogunov, A.; Umari, P.; Wentzcovitch, R. M. *Journal of Physics: Condensed Matter* **2009**, *21*, 395502.

(22) Smogunov, A.; Dal Corso, A.; Tosatti, E. *Phys. Rev. B* **2008**, *78*, 014423.

(23) Dal Corso, A.; Mosca Conte, A. *Phys. Rev. B* **2005**, *71*, 115106.

(24) Perdew, J. P.; Burke, K.; Ernzerhof, M. *Phys. Rev. Lett.* **1996**, *77*, 3865–3868.

(25) Bozzolo, G.; Ferrante, J. *Phys. Rev. B* **1992**, *46*, 8600–8602.

(26) Blum, V.; Gehrke, R.; Hanke, F.; Havu, P.; Havu, V.; Ren, X.; Reuter, K.; Scheffler, M. *Computer Physics Communications* **2009**, *180*, 2175–2196.